\documentclass[prd,aps,preprintnumbers,showpacs]{revtex4}

\usepackage{epsfig}
\usepackage{graphicx}
\usepackage{bm}
\usepackage{mathrsfs}

\newcommand{\be}{\begin{equation}}
\newcommand{\ee}{\end{equation}}
\newcommand{\bea}{\begin{eqnarray}}
\newcommand{\eea}{\end{eqnarray}}
\newcommand{\nn}{\nonumber}

\def\a{\alpha}
\def\b{\beta}
      \def\D{\Delta}

        \def\F{\Phi}

\def\g{\gamma}      \def\G{\Gamma}

\def\l{\lambda}   \def\L{\Lambda}
\def\m{\mu}

\def\p{\pi}      
   
\def\r{\rho}

  \def\U{\Upsilon}

\begin{document}


\title{Final stare interaction enhancement effect on the near threshold
$p\bar p$ system in $B^\pm\to p\bar p \p^\pm$ decay}

\author{\textbf{Vincenzo Laporta}}
\email{vincenzo.laporta@cpht.polytechnique.fr}
\affiliation{Centre de Physique Th{\'e}orique, {\'E}cole Polytechnique, CNRS, 91128 Palaiseau Cedex, France.}

\begin{abstract}
We discuss the low-mass enhancement effect in the baryon-antibaryon
invariant mass in three-body baryonic $B$ decays using
final state interactions in the framework of Regge theory. We show that the rescattering
between baryonic pair can reproduce the observed mass spectrum.
\end{abstract}

\maketitle

\section{Introduction}

The mesonic $B$ decays have been intensively studied principally to test Standard Model 
and to investigate the $CP$ violation mechanism as predict by the Cabibbo-Kobayashi-Maskawa model. 
On the other hand, the large mass of the $b$-quark allows $B$ meson to decay into a 
pair baryon--antibaryon too. The first measurements of baryonic $B$ decays were 
reported by ARGUS collaboration \cite{Albrecht:1988gd}, at the end of '80 years, and 
stimulated extensive theoretical studies. Interest in this area was revitalized in the last 
years thanks to the new measurements collected by CLEO, BELLE and \textsc{BaBar}. 
The phenomenology of baryonic $B$ decays is rich and diversified (for recent reviews see 
Ref.\cite{Kichimi:2004zu,Cheng:2003fq,Cheng:2006nm}): in this paper we will refer to three-body $B$ decays with a 
baryon-antibaryon pair and a meson in the final state.

\begin{figure}[b!]
\includegraphics[scale=.4]{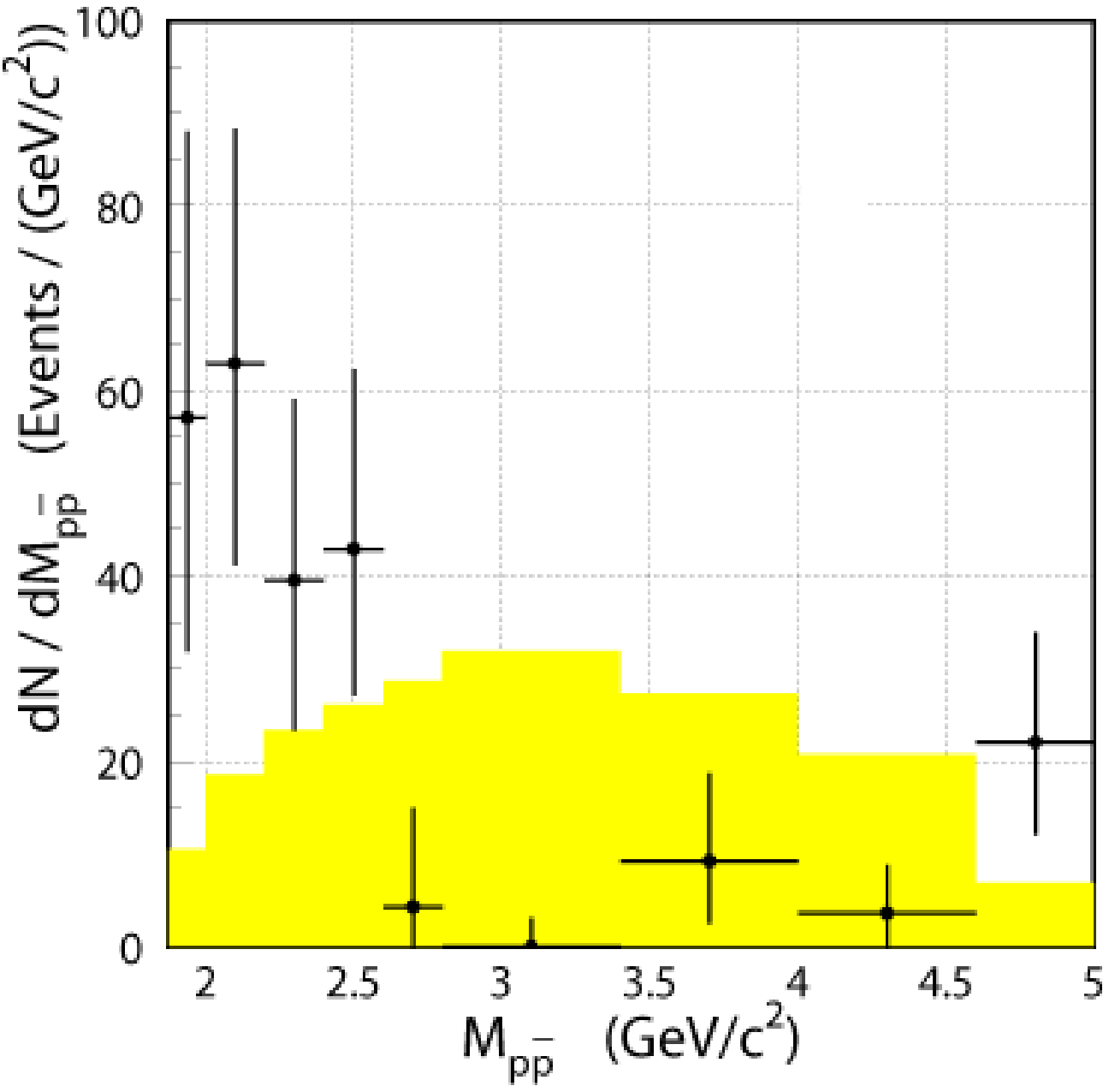}\hspace{2cm}
\includegraphics[scale=.4]{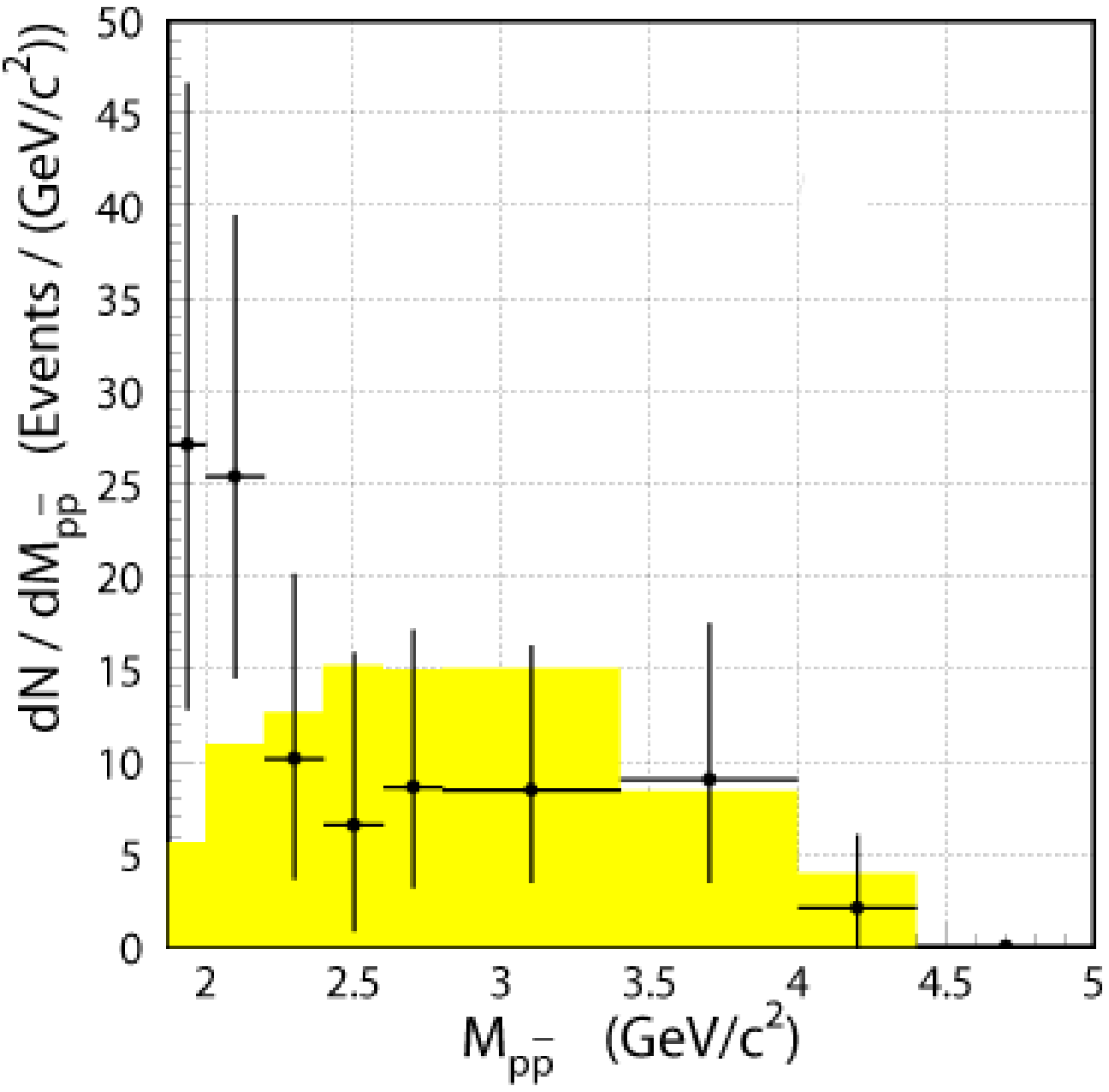}
\caption{Experimental results for $B^+\to p\bar p\pi^+$ and $B^+\to p\bar pK^{*+}$ respectively on the left and right panel. The shared distribution is from the phase-space MC simulation with area normalized to signal yield \cite{Wang:2003iz}. In both cases is evident the sharp peak near to $m_{p\bar p}\approx 2m_p\approx 2$ GeV. \label{fig.exp}}
\end{figure}

There is a common and unique feature for baryon--antibaryon--meson $B$ decays, to which this 
paper is devoted, that is the observed peak, near to the threshold area, in the invariant 
mass spectrum of baryon-antibaryon system. The first experimental observation of this enhancement 
came from Belle collaboration studying the proton-antiproton system for the decays 
$B\to p\bar p K$ \cite{Abe:2002ds} and $B\to p\bar pD^{(*)}$ \cite{Abe:2002tw}. 
Very recently \textsc{BaBar} collaboration confirmed the threshold enhancement in the 
same channels \cite{Aubert:2005gw,Aubert:2006qx} with a very high statistics. 
Next the same peak has been found by BELLE studying the channels 
$B\to p\bar p\pi$ and $B\to p\bar p K^*$ \cite{Wang:2003iz}: these results are reported in Fig. \ref{fig.exp}.

The low-mass enhancement effect in three-body baryonic $B$ decays indicates,
thus, that there is a favorable experimental configuration,
the  baryon-antibaryon pair with low invariant mass accompanied
by a fast recoil meson, that is preferred respect to the other. Therefore,
the main consequence of the threshold effect is that the two-body baryonic decays,
which have an invariant mass fixed to $m_B$, should be disfavored respect to three-body decays.
In fact it has been well established experimentally that 
\be \label{eq.Br3to2} {\cal B}(B\to {\cal B} \,\bar{\cal B}' M)\gg {\cal B}(B\to {\cal B}\,\bar{\cal B}')\,; 
\ee typically ${\cal B}(B\to {\cal B}\,\bar{\cal B}' M)\approx{\cal O}(10^{-6})$ 
and ${\cal B}(B\to {\cal B}\,\bar{\cal B}')\approx{\cal O}(10^{-7})$ \cite{Kichimi:2004zu,Cheng:2003fq,Cheng:2006nm}. 
In other words, thanks to the threshold effect, in $B\to {\cal B}\,\bar{\cal B}' M$ decays 
the effective mass of baryon-antibaryon is reduced, respect to the baryonic two-body decays, 
as the emitted meson carries away much energy. The hierarchy on the Branching ratio 
in (\ref{eq.Br3to2}) is thus an indirect support to the enhancement effect.

Several theoretical speculations have been proposed to interpret the anomalous observed 
peak at the threshold. The first discussion was given by Hou and Soni using a 
simple pole model \cite{Hou:2000bz}, and subsequently this mechanism 
has been addressed in the QCD naive-factorization approach \cite{Cheng:2001tr,Chua:2002wn}. 
Various interpretations include models with baryon-antibaryon bound state based on 
the old idea of Fermi and Yang \cite{Fermi:1959sa}, while exotic interpretations are 
given in terms of glueball intermediate states and fragmentation picture as in 
Ref. \cite{Rosner:2003bm}. Furthermore, the problem has been studied also in 
the framework of perturbative QCD: in this scheme the enhancement is led back to 
the asymptotic behaviour of baryonic form factors \cite{Geng:2006wz}.

In this paper we will follow a different philosophy. The threshold enhancement
effect in proton-antiproton invariant mass, is not an exclusive prerogative of
$B$ meson decay but it has been observed also in other meson decays, with a very
similar behavior: in the $J/\psi\to\g p\bar p$ decay by BES collaboration \cite{Bai:2003sw}
and from CLEO measurements in $\U(1S)\to\g p\bar p$ decay \cite{Athar:2005nu}.
Furthermore, recent observations made at the LEAR at CERN, studying the
proton-antiproton production in the cross section of the reaction $e^+e^-\to p\bar p$
\cite{Klempt:2002ap}, showed a near threshold structure similar to the previous one.
Moreover the characteristic near threshold enhancement is not only present in the
proton-antiproton system but also in the systems that contain hyperons as $\L \bar p$
and $\bar\L p$, as observed by BELLE in $B^0\to p\bar\L\p^-$ decay \cite{Wang:2005fc}
and by BES in $J/\psi\to p\bar \L K^-+c.c.$ decay \cite{Ablikim:2004dj}, and
in the system $\L\bar\L$ as observed by BELLE in $B^+\to\L\bar\L K^+$ decay \cite{Lee:2004mg}.

As one can see from the list of experimental data, the enhancement effect
is universal and it is present with similar feature in different contexts:
$B$ decays as well as $J/\psi$ and $\U$ decays, it is found in the $p\bar p$ system as well
as in $\L \bar p$ system etc$\ldots$. Therefore these experimental results suggest
that: \begin{quote}\emph{the low-energy enhancement in baryon-antibaryon mass spectrum is
mostly related with the dynamics of the baryonic pair and is weakly depending by the
production decay vertex.}\end{quote}
For this reason we consider the threshold enhancement effect
in decays as $B\to {\cal B}\,\bar {\cal B}' M$ is due to the
interactions in the baryon-antibaryon system.
We will return on this hypothesis in Section \ref{sec.discuss}.
Rescattering effect have been studied, in $B$ decays as well as
in $J/\psi$ decays, in some recent papers based on potential-like model
for baryon-antibaryon interaction (Paris or Bonn potential) or by using
one--pion--exchange model (OPE) \cite{Entem:2007bb,Haidenbauer:2006au,Sibirtsev:2004id}.
In this paper we address the final state interactions in the
framework of Regge theory, in particular we will study in detail the channel $B$ into $p\bar p\p$.

\section{Final state interactions for three-body decays \label{sec.Smatr}}

\begin{figure}[b!]
\includegraphics[scale=.45]{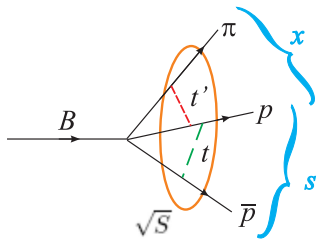}
\caption{Final state interactions for three-body decay. Definition of kinematic variables.\label{fig.rescatt}}
\end{figure}
\begin{figure}[t!]
\includegraphics[scale=.45]{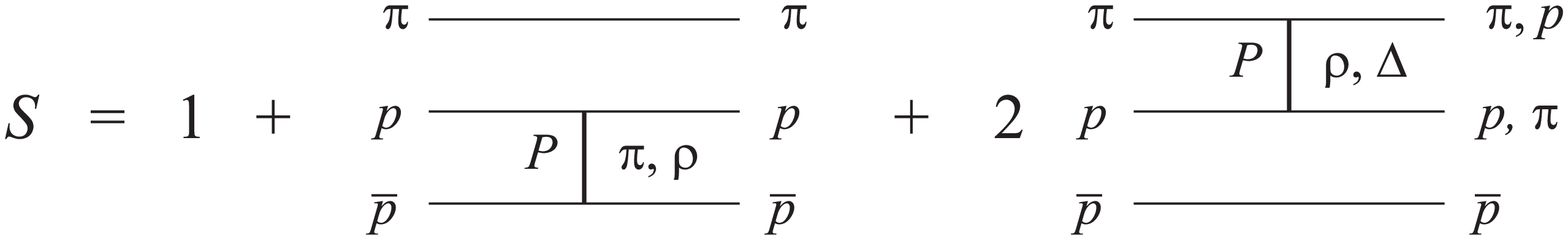}
\caption{Diagrams that contribute to the rescattering. $\cal P$ is the Pomeron and $\p,\r,\D$ are Regge trajectories.\label{fig.Smatr}}
\end{figure}

In this section we write some basic formula for final state interactions model, in the formalism of Regge theory \cite{Donoghue:1996hz}, for three-body decays.
In particular we study in detail $B\to p\bar p\p$. We start to define the kinematic variables as
\be s=(p_p+p_{\bar p})^2\,, \hspace{2cm} x=(p_p+q)^2\,, \hspace{2cm} y=(p_{\bar p}+q)^2\,, \ee
respectively the invariant mass of proton-antiproton, proton-pion and antiproton-pion system;
they satisfy the constraint $s+x+y=m_B+2m_p+m_\p$.
In the following we neglect the pion mass respect to the other ones and we
work with $s$ and $x$ as independent variables. The relation between the `full' amplitude $\cal A$,
where the final state interactions are take into account,
and the `bare' amplitude $\tilde {\cal A}$ is given by the Watson--Migdal
theorem \cite{Watson:1952ji}: \be {\cal A}=\sqrt {\cal S} \tilde{\cal A} \,, \ee
where $\cal S$ is the rescattering strong interaction $S$-matrix
for a given partial wave ($J=0$ in the present case).
In the following we consider rescattering without flavour changing,
that is our intermediate state is the same of the outgoing one.
Moreover, at leading order, for completeness one should consider the interaction between particles
in all possible way as shown in Fig. \ref{fig.rescatt}.
The $S$-matrix (with the initial state = final state) can be
written in the following way \cite{Collins:1977jy}
\be {\cal S}=1+2i\frac{\sqrt{\l(s,m_p^2,m_p^2)}}{s}\, T^{p\bar p}(s)+2i\frac{\sqrt{\l(x,m_p^2,m_\p^2)}}{x}\, T^{p\p}(x)+2i\frac{\sqrt{\l(y,m_p^2,m_\p^2)}}{y}\, T^{\bar p\p}(y)\,,\label{eq.Smatr} \ee where $\l(s,m_p^2,m_p^2)$ is the Kall\'{e}n function and the $T$-matrices contains the elastic and also inelastic rescattering effects. From symmetry considerations we have $T^{p\pi}=T^{\bar p\pi}$. The contributions to the $T$-matrices are shown in Fig. \ref{fig.Smatr}.

The leading contribution to the $T$-matrices, the Pomeron, can be
parameterized, for small and negative transferred momentum $t$,
in a universal form as \cite{Donnachie:1992ny}
\be \label{eq.pomeron} {\cal P}(s,t) = -\b(t)
\left(\frac s{s_0}\right)^{\a_{\cal P}(t)} e^{-i\frac\p2 \a_{\cal P}(t)}\,, \ee
with $s_0\approx {\cal O}$(1 GeV$^2$) and $\a_{\cal P}(t)=1.08+0.25\,t$;
this form is suggested by fit to hadron-hadron scattering total cross sections.
The prefactor $\b(t)$ in Equation (\ref{eq.pomeron}) represents the Pomeron
residue, that is Pomeron `coupling' to the hadrons; phenomenologically
it is given by \cite{Donoghue:1996hz,Zheng:1995mg}
\be \b(t)=\b^{\cal P}\,\frac1{(1-t/0.71)^2}\approx\b^{\cal P}\,e^{2.82t}\,. \ee
The coefficient $\b^{\cal P}$ writes in term of $\b^{\cal P}_{uu}$,
the ${\cal P}-$quark coupling (we consider $SU(2)$ symmetry),
and applying the additive quark counting rule \cite{Donnachie:1992ny}
at each vertex; for instance for the first and the second diagram
in Fig. \ref{fig.Smatr} we have respectively:
\be \b^{\cal P}_{p\bar p}=\left(3\,\b^{\cal P}_{uu}\right)^2\,,
\hspace{3cm} \b^{\cal P}_{p\p}=\left(3\,\b^{\cal P}_{uu}\right)\left(2\,\b^{\cal P}_{uu}\right)\,, \ee
where $\b^{\cal P}_{uu}=1.87$ GeV$^{-1}$ \cite{Zheng:1995mg}.

The non-leading contributions to the rescattering are given by Regge trajectories.
These amplitudes can be parameterized by means the Veneziano formula
\cite{Veneziano:1968yb,Irving:1977ea} in the following way \be \label{eq.regge}
{\cal R}(s,t)=-\b^{\cal R}\,\frac{1+(-1)^{s_{\cal R}}
e^{-i\p\a_{\cal R}(t)}}{2}\,\G\left(\ell_{\cal R}-\a_{\cal R}(t)\right)\,
(\a')^{1-\ell_{\cal R}}\,(\a's)^{\a_{\cal R}(t)}\,, \ee
where $\G$ is the Euler function and \be \a_{\cal R}(t)=s_{\cal R}+\a'(t-m_{\cal R}^2)\,, \ee
is the Regge trajectory with the universal slope $\a'=0.93$ GeV$^{-2}$.
In Table \ref{tab.reggepar} we list the parameters $s_{\cal R}$, $\ell_{\cal R}$
and $m_{\cal R}$ for the trajectories $\p$, $\r$ and $\D$. $\b^{\cal R}$
in Eq.(\ref{eq.regge}) is the residue for the Regge trajectory and,
as in the case of Pomeron, it factorizes in a product of two
residue one for each diagram vertex. To estimate the Regge residue
we note that near $t\approx m_{\cal R}^2$, Eq.(\ref{eq.regge})
reduces to a Feynman-like amplitude \cite{Ladisa:2004bp}
\be  {\cal R}(s,t)\approx -\b^{\cal R}\,\frac{s^{s_{\cal R}}}{t-m_{\cal R}^2}\,, \ee
which allows us to identify $\b^{\cal R}$ as a product of coupling constants, one for each vertex.
\begin{table}
\begin{center}
\begin{tabular}{cccc}
\hline\hline ~Trajectory ${\cal R}$~ & ~~$s_{\cal R}$~~ & ~~$\ell_{\cal R}$~~ & ~~~$m_{\cal R}$(MeV) \cite{Yao:2006px}~~~
\\
\hline
$\pi$ & $0$ & $0$ & $139$
\\
$\rho$ & $1$ & $1$ &$775$
\\
$\D$ & $3/2$ & $3/2$ & $1232$
\\
\hline\hline
\end{tabular}
\caption{Parameters of the Regge trajectories.\label{tab.reggepar}}
\end{center}
\end{table}

To estimate the $\D$--proton--pion coupling we used  an effective lagrangian approach with the experimental value of $\D\to p\p$ decay width. The interaction lagrangian can be written as \be \mathscr{L}_{\D p\p}=g_{\D p\p}\bar\psi\,\chi_\m\,\partial^\m\p\,+\,c.c.\,, \label{eq.deltacoupling} \ee where $\psi$ is the standard fermionic field for the proton, $\p$ the pion field and $\chi_\m$ the Rarita-Schwinger tensor \cite{Rarita:1941mf} for a $\frac32$-spin field that describes the $\D$. From Lagrangian in Eq.(\ref{eq.deltacoupling}) one can calculate the following decay width \be \G(\D\to p\p)=\frac1{8\p}\,p_f^3\,g_{\D p\p}^2\,\frac{(m_\D+m_p)^2-m_\p^2}{3\,m_\D^2}\,, \ee and by means his experimental value, $\G_{exp}(\D\to p\p)=118$ MeV at $p_f=229$ MeV \cite{Yao:2006px}, one can determinate the coupling constant \be g_{\D p\p}=15.4\;\; \textrm{GeV}^{-1}\,. \ee In the following we will neglect the $t$-dependence of $\b^\D_{p\p}$ Regge residue and we will identify it as $g_{\D p\pi}$. The other residues useful in our treatment are given by \cite{Irving:1977ea} \be \b^\p_{pp}=25.2\,,\hspace{2cm} \b^\r_{pp}=13.0\,, \hspace{2cm} \b^\r_{\p\p}=-8.0\,. \ee

Therefore the $T$-matrices present in Eq.(\ref{eq.Smatr}) write as \cite{Byckling:1969sx,Zheng:1995mg} \bea T^{p\bar p}(s)&=&\frac1{16\p}\frac s{\l(s,m_p^2,m_p^2)} \int_{-s+4m_p^2}^0\ {\cal P}_{p\bar p}(s,t)+{\cal R}_{p\bar p}^\pi(s,t)+{\cal R}_{p\bar p}^\r(s,t)\ dt \label{eq.Tpp}
\\
T^{p\p}(x)&=&\frac1{16\p}\frac x{\l(x,m_p^2,m_\p^2)} \int_{-\frac{(m_p^2-x)^2}{x}}^0\ {\cal P}_{p\p}(x,t)+{\cal R}_{p\p}^\r(x,t)+{\cal R}_{p\p}^\D(x,t)\ dt\,,\label{eq.Tppi} \eea and finally the double differential width for $B\to\p p\bar p$, with the final state interaction effect take into account, is thus given by \cite{Byckling:1969sx} \bea d\G' &=&\frac1{(2\p)^3} \frac1{32 m_B^3} |{\cal A}|^2\, dx\, ds \nn
\\
&=& \frac1{(2\p)^3} \frac1{32 m_B^3} |{\cal S}|\,|\tilde{\cal A}|^2\, dx\, ds\,.\label{eq.2width} \eea

\section{Discussion and results \label{sec.discuss}}

The double differential width writes in Eq.(\ref{eq.2width}) depends on two functions: (i) The bare amplitude $\tilde{\cal A}$ that describes the $B$ decay vertex and (ii) the $S$-matrix ${\cal S}$ that describes the re-interaction between the outgoing particles. In Section \ref{sec.Smatr} we discussed the rescattering contributions; the problem now is to understand the role that plays the bare amplitude in the full dynamics. How we saw in the Introduction experimental data show that the characteristic peak at the threshold in the proton-antiproton system as well as in other different baryonic systems has a very universal beaviour. On the other hand the bare amplitude should be strongly dependent from the dynamics of weak decay vertex, or in other words, it should be different for different final states. While the strong interactions are independent from dynamics of particle productions and they are very similar beaviour between hadrons. Therefore, how previously argued in the Introduction, one should realize that the dynamics of the process is principally dominated by rescattering interactions between baryons and $B$ decay vertex plays a `fine-tuning' role on the shape of differential width.

In summary we neglect the variations of bare amplitude respect to $S$-matrix
and we trait $\tilde{\cal A}$ as was a constant. Thus if we turn-off the
rescattering effects (${\cal S}=1$) and we integrate-out the $x$ variable in Eq.(\ref{eq.2width}),
one obtains a differential width proportional to three-body phase-space $\F(s)$, namely
\be \F(s)=\frac1{2^8\p^3m_B^3}\frac{m_B^2-s}{\sqrt s}\sqrt{s-4m_p^2}\,, \ee
that the bare amplitude can only modulate.

\begin{figure}
\includegraphics[scale=1]{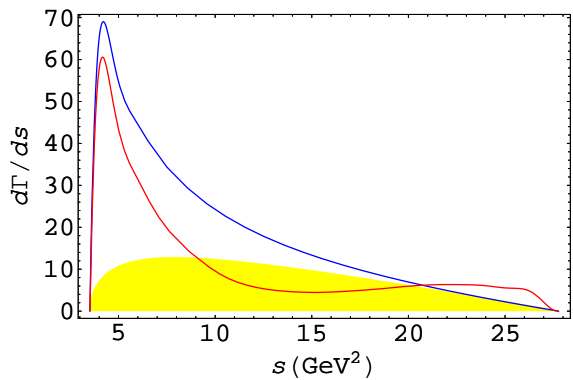}
\caption{Differential width of $B\to p\bar p\p$ as function of invariant mass of proton-antiproton. Shared yellow region is just the phase-space $\F(s)$. Blue line refers to the final state interactions effect with only re-interaction in the proton-antiproton system ($T^{p\bar p}$), while red line take into account full rescattering ($T^{p\bar p}+2\,T^{p\p}$). Units are arbitrary.\label{fig.plotfull}}
\end{figure}

\begin{figure}
\includegraphics[scale=1]{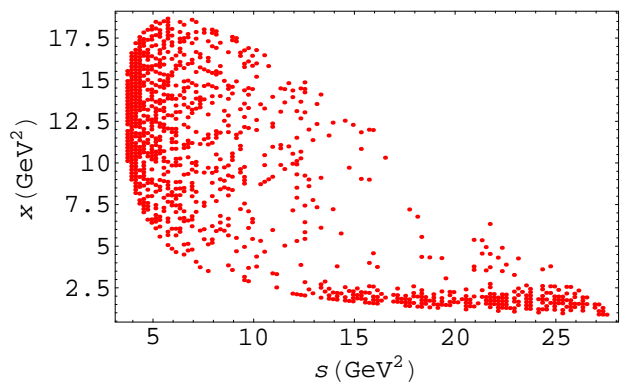}
\caption{Dalitz plot of $B\to p\bar p\p$ with re-interaction effects. We can distinguish two regions in the diagram: First on the left due to the Pomeron exchanged in proton-antiproton system and one at bottom on the right that is the exchange of the $\D$ between pion and (anti)proton. \label{fig.dalitz}}
\end{figure}

The effects of final state interactions on the differential width of $B\to p\bar p\p$ (modulo $|\tilde{\cal A}|^2$) are reported in Fig. \ref{fig.plotfull} where the blue line takes into account only the effects of re-interaction between proton-antiproton (see Eq.(\ref{eq.Tpp})) while red curve contains the rescattering with the pion too (Eqq.(\ref{eq.Tpp}) and (\ref{eq.Tppi})). There are two characteristics that emerge from full rescattering in Fig. \ref{fig.plotfull}: (i) The strong peak near to the threshold area and (ii) the `see-saw' trend for higher values of $s$. Both are present in experimental data, compare for instance Fig. \ref{fig.exp}.  These features are better understood if we look at the Dalitz plot in Fig. \ref{fig.dalitz}; how one can see from the structure of the Dalitz distribution, respect to the $s$ variable, the peak is essentially due to the Pomeron for rescattering between the baryonic pair (structure on the left of plot) while the latter is due to the presence of the Pomeron and $\D$ in the system pion-(anti)proton (structure on the bottom).

\begin{figure}
$B\to p\bar pK$ \hspace{4cm} $B\to p\bar pK^*$ \hspace{4cm} $B\to p\bar pD$

\includegraphics[scale=.7]{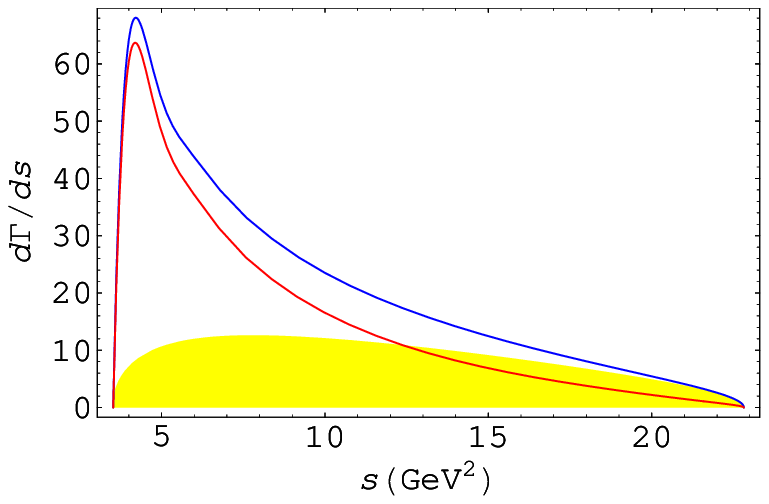}
\includegraphics[scale=.7]{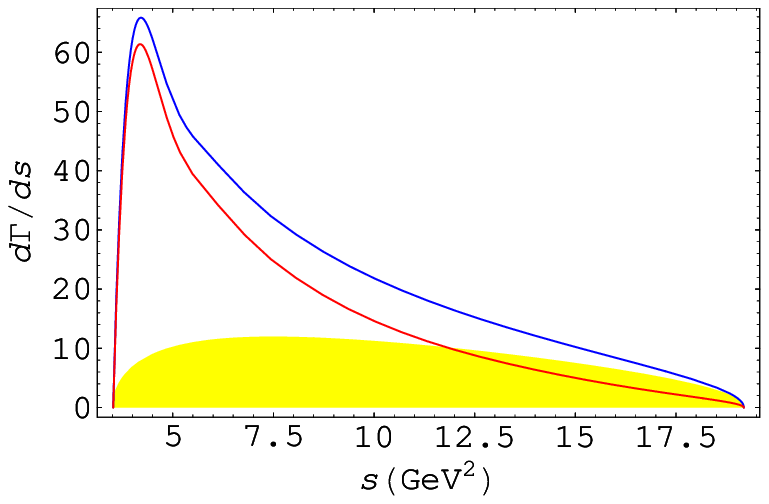}
\includegraphics[scale=.7]{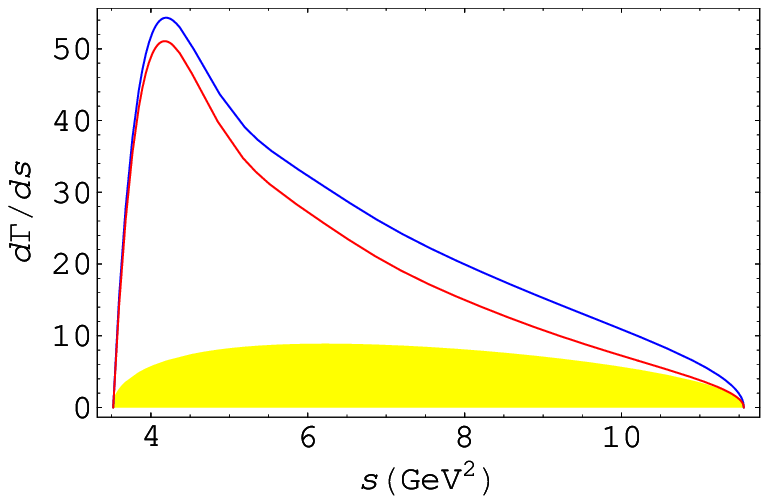}
\caption{Differential width (left to right) of $B\to p\bar pK$,
$B\to p\bar pK^*$ and $B\to p\bar pD$ as function of invariant
mass of proton-antiproton. Shared yellow region is just the phase-space.
Rescattering contributions in proton-antiproton system are the same of Fig.
\ref{fig.plotfull} (blue line), while for full rescattering (red line) the
systems $pK$, $pK^*$ and $pD$ contain only Pomeron contribution.
Units are arbitrary.\label{fig.ppbarX}}
\end{figure}

To further test our analysis we applied rescattering approach to the
channels $B\to p\bar pK$, $B\to p\bar pK^*$ and $B\to p\bar pD$.
In these cases for the systems $pK$, $pK^*$ and $pD$ we take into
account only the Pomeron contribution; results are in Fig. \ref{fig.ppbarX}.
How one can see also in these cases the rescattering model is able to reproduce
the peak near to the threshold area but in some cases, as in the channel $B\to p\bar pK^*$
for instance, the spectrum is not well reproduced for higher $s$: Probably in this case
one should add another rescattering contributions between $K^*$ and (anti)proton, but the
peak is, in any case, due to the rescattering between baryonic pair. Moreover the shape of
spectrum is similar in the three cases, but as the phase-space of $s$ variable become smaller
(\emph{i.e.} the outgoing mass of meson increase) the peak width tend to enlarge.

\begin{figure}
$B\to J/\psi p\bar p$ \hspace{5.5cm} $B\to J/\psi \L \bar p$

\includegraphics[scale=.7]{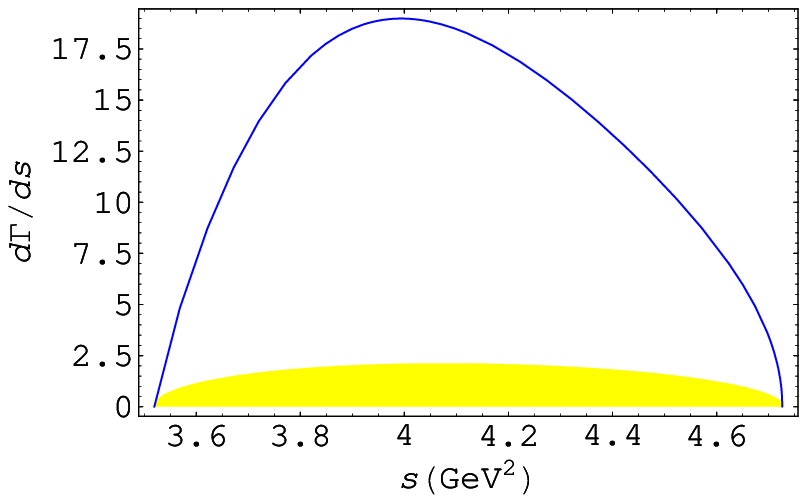}\hspace{2cm}
\includegraphics[scale=.7]{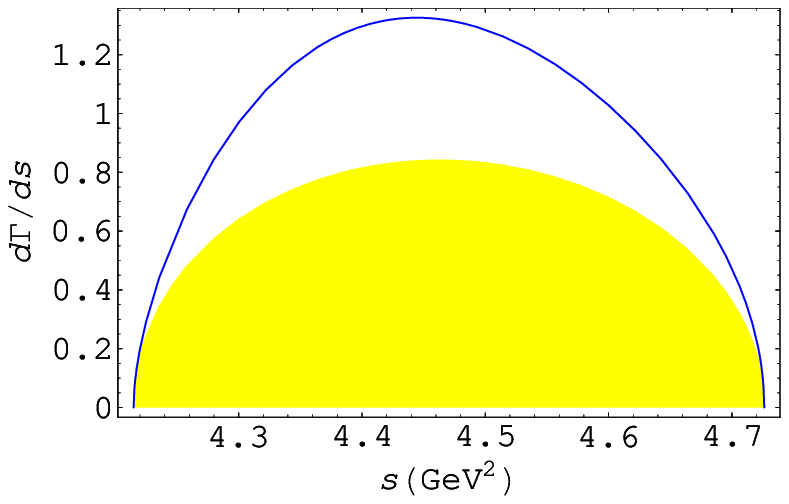}
\caption{Differential width (left to right) of $B\to J/\psi p\bar p$
and $B\to J/\psi\L \bar p$ as function of invariant mass of baryon-antibaryon.
Shared yellow region is just the phase-space. Rescattering contributions are
only in baryon-antibaryon system. Units are arbitrary.\label{fig.ppbarJpsi}}
\end{figure}

There is another support to our thesis of the re-interaction, and it
comes again from experimental data. In fact recently BELLE studied $B$
decay channels with an heavy particle in the final state, precisely $B\to J/\psi \L\bar p$
and $B\to J/\psi p\bar p$ \cite{Xie:2005tf}. From these measurements emerged that there
is not threshold enhancement effect at all in the invariant mass of $\L \bar p$
when there is a heavy meson in the final state (on the contrary the peak return
if meson is light one as in the case $B\to \L\bar p\p$). Our results for
$B\to J/\psi \L\bar p$ and $B\to J/\psi p\bar p$ are reported in Fig. \ref{fig.ppbarJpsi}:
For these cases we consider only the rescattering between baryon-antibaryon system because
the phase-space of $x$ variable is small and it gives a negligible contribution to
the full amplitude; moreover in the system antiproton-$\L$ we consider the contribution
of Pomeron and the Regge trajectory of $\r$. These two channels are the limit cases
of $B\to p\bar pD$ where the phase-space of $s$ is so small that only the peak is present.

\section{Conclusions \label{sec.concl}}

In this paper we presented a simple model to understand the threshold enhancement effect present in the spectrum of many baryonic three-body $B$ (and $J/\psi$) decays. This model is based on hypothesis of re-interactions between outgoing particle. We studied in detail the $B\to p\bar p \p$ channel because in this case we are able to make a full analysis of main rescattering diagrams in the proton-antiproton system as well as in the pion-(anti)proton system, but the results can be easily extend to other decay channels. Our principal ansatz is the constance of bare amplitude, or better, its variations from a channel decay to another are negligible respect to the rescattering effects. This feature is suggest from experimental behaviour.

In this framework we are able to reproduce the experimental
enhancement effect near to the threshold area for $B\to p\bar p\p$
decay (see Fig. \ref{fig.plotfull}) and with some change also in other
decays where the pion is substituted by a heavier meson (see Fig. \ref{fig.ppbarX}).
We studied also two channels with a heavy meson in the final state, precisely
$B\to J/\psi p\bar p$ and $B\to J/\psi\L \bar p$ (see Fig. \ref{fig.ppbarJpsi}):
in these cases due to the phase-space configuration the peak disappear, or better
it span all spectrum. This feature has been experimentally observed in $B\to J/\psi\L \bar p$
decay; on the other hand, for $B\to J/\psi p\bar p$ it is a theoretical prediction.

\acknowledgments{I wish to thank Prof. B. Pire and Dr. M. Ruggieri
to careful reading of this manuscript, and Dr. A.D. Polosa and Dr.ssa F. Pagliarulo
for encouragement.}

\end{document}